\documentclass[11pt]{article}

\usepackage{a4wide}
\usepackage{amsmath,amssymb,amsthm,amscd}
\usepackage{graphicx}
\usepackage{authblk}

\usepackage{natbib}
\usepackage{algorithm}	
\usepackage{algorithmic}
\usepackage{bm}
\usepackage{bbm}
\usepackage{enumitem}

\usepackage{listings}
\usepackage{setspace}

\usepackage[toc,page]{appendix}

\theoremstyle{plain}

\providecommand{\keywords}[1]{\textbf{\textit{Keywords: }} #1}

\DeclareMathOperator{\plim}{plim}

\begin{document}

\title{Langevin Incremental Mixture Importance Sampling}

\author[1,$\dag$]{Matteo Fasiolo}
\author[2]{Fl\'avio Eler de Melo}
\author[2]{Simon Maskell} 
\affil[1]{School of Mathematics, University of Bristol, United Kingdom.}
\affil[2]{Department of Electrical Engineering and Electronics, University of Liverpool, Uniter Kingdom.}
\affil[$\dag$]{Correspondence: matteo.fasiolo@bristol.ac.uk}

\maketitle

\abstract{
This work proposes a novel method through which local information about the target density can be used to construct an efficient importance sampler. The backbone of the proposed method is the Incremental Mixture Importance Sampling (IMIS) algorithm of \cite{raftery2010estimating}, which builds a mixture importance distribution incrementally, by positioning new mixture components where the importance density lacks mass, relative to the target. The key innovation proposed here is that the mixture components used by IMIS are local approximations to the target density. In particular, their mean vectors and covariance matrices are constructed by numerically solving certain differential equations, whose solution depends on the gradient field of the target log-density. The new sampler has a number of advantages: a) it provides an extremely parsimonious parametrization of the mixture importance density, whose configuration effectively depends only on the shape of the target and on a single free parameter representing pseudo-time; b) it scales well with the dimensionality of the target; c) it can deal with targets that are not log-concave. The performance of the proposed approach is demonstrated on a synthetic non-Gaussian multimodal density, defined on up to eighty dimensions, and on a Bayesian logistic regression model, using the Sonar dataset. The \verb|Julia| code implementing the importance sampler proposed here can be found at https:\//\/github.com\//mfasiolo\//LIMIS.
}

\vspace{4pt}




\section{Introduction}

The efficiency gains brought about by taking into account local information about the target density have been amply demonstrated in the context of Markov chain Monte Carlo (MCMC) sampling. For instance, the seminal paper of \cite{girolami2011riemann} introduced variations of the Metropolis adjusted Langevin (MALA) \citep{roberts1996exponential} and Hamiltonian Monte Carlo (HMC) \citep{duane1987hybrid} samplers which, by exploiting second order information, can efficiently sample highly dimensional non-Gaussian targets. This is achieved using an adaptive proposal, based on the local information contained in the gradient and Hessian of the target log-density. Notably, the state-of-the-art probabilistic programming language \verb|Stan| \citep{carpenter2016stan}, uses the tuned HMC algorithm proposed by \cite{hoffman2014no} as its default sampler. This demonstrates that these ideas have changed MCMC sampling practice as well as theory. It is therefore surprising that these concepts have not been exploited nearly as widely in the context of Importance Sampling (IS).

In this paper we attempt to fill this gap, by extracting local information about the target density and using it to set up an efficient importance sampler. We accomplish this by considering ideas related to Langevin diffusions and adapting them to the context of IS. In particular, we demonstrate how linearized solutions to Langevin diffusions can produce Gaussian densities that often represent accurate local approximations to the target density. These local densities can then be combined to form a global mixture importance density that closely approximates the target. To achieve this, we exploit the Incremental Mixture
Importance Sampling (IMIS) algorithm, originally proposed by \cite{raftery2010estimating}. This is an automatic and non-parametric approach to IS, which constructs a mixture importance density by iteratively adding mixture components in areas where the importance density lacks mass relative to the target. As the examples will demonstrate, the proposed modification of the IMIS algorithm leads to a scalable and semi-automated approach to Importance Sampling (IS). 

The literature related to the current proposal is quite sparse. Indeed, the use of local target information has been adopted mostly in the context of Sequential Monte Carlo (SMC) samplers and particle filtering, rather than IS itself\footnote{Note that, of course, most of the conventional SMC samplers and particle filters are based on IS.}. In particular, \cite{sim2012information} and \cite{schuster2015gradient} consider using MALA's adaptive proposal within SMC samplers. These proposals are quite different from our approach, because we iteratively construct a single mixture importance density, not a sequence of them. In addition, in our proposal the mean and covariance of the mixture components are not based on the derivatives of the target log-density at a single fixed location, as in MALA, but are obtained by numerically integrating certain differential equations, whose solution depends on the shape of whole regions of the target. Also, while in SMC each sample is generally perturbed individually, in our case the number of mixture components is much lower than the number of samples, which reduces the cost of constructing the importance distribution and of evaluating its density.

In the context of particle filtering, \cite{bunch2015approximations} propose a Gaussian particle flow method, which aims at approximating the optimal importance density of a class of non-linear Gaussian state space models. In particle flow algorithms \citep{daum2008particle} a particle is moved continuously in pseudo-time according to differential equations that depend on the underlying shape of the target density. The drawback of many particle flow algorithms is that, despite their theoretical elegance, implementing them for general models requires several layers of approximation, whose effect is not easy to quantify \citep{bunch2015approximations}. Even though we are not considering particle filtering here, our current work has been inspired by this literature. A critical distinguishing characteristic of our proposal is that we exploit local information about the target density, while not introducing any extra approximation or source of bias in the importance sampler. 

The rest of the paper is structured as follows. In Section \ref{sec:IMIS} we briefly describe the IMIS algorithm of \cite{raftery2010estimating}. Then, in Section \ref{sec:linearLangevin}, we show how the solutions to linearized Langevin diffusions can be used to generate local approximations to the target density and we explain how these can be exploited within the IMIS algorithm. This results in the new Langevin IMIS (LIMIS) sampler. Calculating the mean vector and covariance matrix of each importance mixture component requires solving certain differential equations. This has to be done numerically, and in Section \ref{sec:stepSizeSelection} we propose a novel statistically-motivated criterion for selecting the step-size of the numerical integrator. In Section \ref{sec:Examples} we compare the new sampler to IMIS, MALA and IS on two examples. The first is a multimodal mixture density, whose components are warped Gaussian densities, defined on up to 80 dimensions. In the second we sample the posterior of Bayesian logistic regression model, using the Sonar dataset of \cite{gorman1988analysis}. Section \ref{sec:computCons} contains some discussion of the computational cost of each method, while Section \ref{sec:TuningT} explains how the pseudo-time of integration can be selected in an automatic fashion. We summarize the results and discuss possible future directions in Section \ref{sec:conclusion}.

\section{Incremental Mixture Importance Sampling} \label{sec:IMIS}

The IMIS algorithm is an automatic and non-parametric approach to IS, which is particularly
useful for highly non-Gaussian target densities \citep{raftery2010estimating}.
Let $\pi({\bf x})$ and $p({\bf x})$ be, respectively, the target and the prior
densities, with ${\bf x}\in \mathbb{R}^{d}$. Here we describe a slightly modified version of IMIS, which includes the following steps:

\subsubsection*{Algorithm 1: Nearest Neighbour IMIS (NIMIS)} 
\begin{enumerate}
\item Initialization:

\begin{enumerate}
\item Sample $n_{0}$ variables, ${\bf x}_{1}, \dots, {\bf x}_{n_0}$, from $p({\bf x})$.
\item Calculate the weight of each sample
\[
w_{i}^{0}=\frac{\pi({\bf x}_{i})}{p({\bf x}_{i})},\;\;\;\text{for}\;\;\;i=1,\dots,n_{0}.
\]
\end{enumerate}
\item Importance Sampling: for $k=1,2,\dots$, repeat

\begin{enumerate}
\item Let ${\bf x}_{j}$ be the sample with the largest weight and define $\bm \mu_{k}={\bf x}_{j}$.
Calculate the covariance, $\bm \Sigma_{k}$, of the $b$ samples
with the shortest Mahalanobis distance from $\bm \mu_{k}$. The metric
used to calculate the distance is the covariance of all the
samples generated so far.
\item Sample $b$ new variables from a multivariate Student's t distribution
with mean $\bm \mu_{k}$, covariance $\bm \Sigma_{k}$ and $\nu>0$ degrees
of freedom.
\item Update the importance weights of all samples generated so far
\begin{equation} \label{eq:wUpdateIMIS}
w_{i}^{k}=\pi({\bf x}_i)\big/\Big\{ \frac{n_{0}}{n_{k}}p({\bf x}_{i})+\frac{b}{n_{k}}\sum_{l=1}^{k}\text{mvt}({\bf x}_{i}|\bm \mu_{l},\bm \Sigma_{l},\nu)\Big\}, \;\;\;\; \text{for} \;\; i=1,\dots,n_{k},
\end{equation}
where $n_{k}=n_{0}+kb$ and $\text{mvt}({\bf x}|\bm \mu,\bm \Sigma,\nu)$ indicates the density of a multivariate Student's t distribution, with location $\bm \mu$, covariance $\bm \Sigma$ and $\nu$ degrees of freedom.
\item If a chosen criterion is met, terminate.
\end{enumerate}
\end{enumerate}
The above algorithm differs from the original IMIS procedure of \cite{raftery2010estimating} in minor respects. In particular, in their version $\bm \Sigma_{k}$ is a weighted covariance, where the $i$-th weight is proportional to $(w_i^k+n_k)/2$. We have verified that these weights can be quite unstable, especially in early iterations and in high dimensions, hence we prefer using an unweighted covariance. In step 2(a) they use the covariance of the prior distribution, rather than the covariance of all the generated samples, to determine the distances. But this approach does not seem appropriate when the prior is not a good approximation to the target. Also, they
also use multivariate Gaussian, rather than Student's t, densities. Our experience suggests that in IS it is better erring on the side of robustness, hence we prefer using Student's t densities to ensure that the proposal is heavier-tailed than the target.

The key idea behind IMIS is that it lets the importance weights determine
where new mixture components should be placed. The fact that the covariance
of the new components is estimated using a Nearest Neighbour approach
is somewhat secondary. For this reason we use the acronym IMIS to refer
to the overall approach, while we use NIMIS to refer to its Nearest Neighbour
version. 

In this work we use ideas related to Langevin diffusions to determine $\bm \mu_{k}$ and
$\bm \Sigma_{k}$ in step 2(a). As we will illustrate empirically in Section \ref{sec:Examples}, this modification is particularly advantageous in high dimensions. However, the purpose of this work not so much improving upon the NIMIS algorithm, but rather showing how local information about the target can be exploited to set up an efficient mixture importance density.

\section{Langevin Incremental Mixture Importance Sampling} \label{sec:linearLangevin}

Consider a $d$-dimensional Langevin diffusion, with stationary distribution $\pi({\bf x})$, which is defined by the stochastic differential equation
\begin{equation}
d{\bf x}_{t}=\frac{d t}{2}\nabla\log\pi({\bf x}_{t})+d{\bf b}_t,\label{eq:langevinSDE}
\end{equation}
where $\nabla\log\pi({\bf x})$ is the gradient of the target log-density and ${\bf b}_t$ is a $d$-dimensional Brownian motion. The dynamics of the first two moments of ${\bf x}_t$ are not available for most target distributions, but if we consider the discrete-time version of (\ref{eq:langevinSDE}), that is
$$
{\bf x}_{t+\delta t} = {\bf x}_{t} + \frac{\delta t}{2}\nabla\log\pi({\bf x}_{t}) + \delta t\hspace{0.1em} {\bf z}_t, \;\;\;\;\; {\bf z}_t \sim N({\bf 0}, {\bf I}),
$$
and we linearize the gradient around $\mathbb{E}({\bf x}_t)$, we obtain
\begin{equation} \label{eq:linDisMean}
\mathbb{E}({\bf x}_{t+\delta t}) \approx \mathbb{E}({\bf x}_t) + \frac{\delta t}{2} \nabla \log\pi \big \{{\mathbb{E}({\bf x}_t)}\big \},
\end{equation} 
and
\begin{equation} \label{eq:linDisCov}
\text{Cov}({\bf x}_{t+\delta t}) \approx \bigg [ {\bf I} + \frac{\delta t}{2} \nabla^{2}\log\pi \big \{{\mathbb{E}({\bf x}_t)}\big \} \bigg ] \text{Cov}({\bf x}_{t}) \bigg [ {\bf I} + \frac{\delta t}{2} \nabla^{2}\log\pi \big \{{\mathbb{E}({\bf x}_t)}\big \} \bigg ]^T + \delta t {\bf I},
\end{equation}
where $\nabla^2\log\pi({\bf x})$ is the Hessian of the target log-density and ${\bf I}$ is a $d$-dimensional identity matrix. In continuous-time this leads to the following differential equations
\begin{equation} \label{eq:diffEqMean}
\dot{\bm \mu}_{t}=\frac{d{\bm \mu}_t}{dt}=\frac{1}{2}\nabla\log\pi(\bm \mu_{t}),
\end{equation}
\begin{equation} \label{eq:diffEqVar}
\dot{\bf \Sigma}_{t}=\frac{d{\bf \Sigma}_t}{dt}=\bigg\{\frac{{1}}{2}\nabla^{2}\log\pi({\bm \mu}_{t})\bigg\}{\bf \Sigma}_{t}+{\bf \Sigma}_{t}\bigg\{\frac{1}{2}\nabla^{2}\log\pi({\bm \mu}_{t})\bigg\}+{\bf I},
\end{equation}
where we defined ${\bm \mu}_{t} = \mathbb{E}({\bf x}_t)$ and ${\bm \Sigma}_{t}= \text{Cov}({\bf x}_t)$. Notice that if the gradient is linear, that is if $\nabla\log\pi({\bf x}) = {\bf F}{\bf x}$ for some matrix $\bf F$, (\ref{eq:diffEqMean}) and (\ref{eq:diffEqVar}) are equivalent to the differential equations used to propagate the mean and covariance of the state process in the Kalman-Bucy filter \citep{bucy1987filtering}, under the special circumstance that the observation and control processes are absent. 

If $\pi({\bf x})$ is Gaussian then $\nabla\log\pi({\bf x})$ is linear and, given any initial state ${\bf x}_{t_0}$, (\ref{eq:diffEqMean}) and (\ref{eq:diffEqVar}) can be solved analytically. In addition, ${\bm \mu}_t$ and ${\bm \Sigma}_t$ will converge, as $t \rightarrow \infty$, to the mean vector and covariance matrix of $\bf x$ under $\pi({\bf x})$. Hence, given that a Gaussian distribution is fully specified by its first two moments, a Gaussian target is recovered exactly. However, if the target is not Gaussian, several issues arise. Firstly (\ref{eq:diffEqMean}) and (\ref{eq:diffEqVar}) generally do not have analytic solutions. This is a relatively mild problem, which can addressed by using a numerical integrator, such as a Runge-Kutta method \citep{ascher1998computer}. 
More importantly, the solutions to (\ref{eq:diffEqMean}) and (\ref{eq:diffEqVar}) will generally not converge to the true mean and covariance under $\pi({\bf x})$, even as $t \rightarrow \infty$. To see this, assume that $\pi({\bf x})$ is unimodal. Given that the solution to (\ref{eq:diffEqMean}) is a steepest ascent curve, $\bm \mu_t$ will eventually converge to the mode of $\pi({\bf x})$. However, unless $\pi({\bf x})$ is symmetric, its mode differs from its mean vector. 

\begin{figure}
\centering{}\includegraphics[scale=0.4]{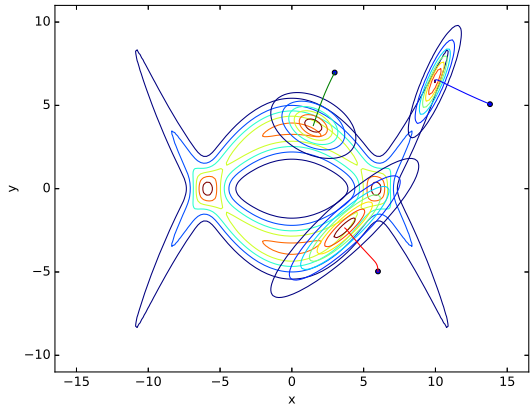}
\caption{Three local Gaussian approximations to a multimodal target density. The mean vectors and covariance matrices of the local densities were generated by solving (\ref{eq:diffEqMean}) and (\ref{eq:diffEqVar}).}\label{fig:linear}
\end{figure}

The second issue entails that, unless $\nabla\log\pi({\bf x})$ is linear, the quality of the approximation to the first two moments will typically degrade as $t-t_{0}$ increases, regardless of the numerical integrator used. This is not of great concern in our case, because we are
interested in creating local, not global, approximations to $\pi({\bf x})$. In particular, let $p({\bf x}_{t_1}|{\bf x}_{t_0})$ be the distribution of ${\bf x}_{t_1}$, generated by integrating  (\ref{eq:langevinSDE}) between $t_0$ and a finite pseudo-time $t_1>t_0$. Also, let $q({\bf x}_{t_1}|{\bf x}_{t_0})$ be a Gaussian approximation to $p({\bf x}_{t_1}|{\bf x}_{t_0})$, with mean and covariance matrix derived by solving (\ref{eq:diffEqMean}) and (\ref{eq:diffEqVar}), with the initial conditions ${\bm \mu}_{t_0} = {\bf x}_{t_0}$ and ${\bf \Sigma}_{t_0} = {\bf 0}$. Our proposal is based on the observation that, while $q({\bf x}_{t_1}|{\bf x}_{t_0})$ might not represent a good global approximation to $\pi({\bf x})$ for any value of $t_1$ or ${\bf x}_{t_0}$, it often provides an accurate local approximation to $\pi({\bf x})$. As an example, consider the highly non-Gaussian density represented in Figure \ref{fig:linear}. In addition to the target, we show three Gaussian densities, obtained by numerically integrating (\ref{eq:diffEqMean}) and (\ref{eq:diffEqVar}) between $t_0=0$ and $t_1=4$, from three starting points. Notice how the covariance matrices adapt to the local shape of the target.

In the context of importance sampling, an accurate global approximation to $\pi({\bf x})$ is needed. We propose to create such a density using a mixture of local Gaussian approximations $q({\bf x}_{t_1}|{\bf x}^1_{t_0}),\dots, q({\bf x}_{t_1}|{\bf x}^k_{t_0})$. The IMIS algorithm provides a natural approach to determining the initial positions, ${\bf x}_{t_0}^1,\dots,{\bf x}_{t_0}^k$, because it places additional mixture components where the importance density is lacking mass, relative to the target. To use the new local linearization within IMIS, it is sufficient to modify step 2(a) of Algorithm 1 as follows:
\begin{description}
\item [{2(a){*}}] Let ${\bf x}_{j}$
be the sample with the largest weight. Given the initial position ${\bm \mu}_{t_{0}} = {\bf x}_{j}$, covariance matrix $\bm \Sigma_{t_0}=\bf 0$
and a user-defined pseudo-time $t_1$, obtain the approximate solutions,
$\hat{\bm \mu}_{t_{1}}$ and $\hat{{\bf \Sigma}}_{t_{1}}$, by numerically integrating (\ref{eq:diffEqMean})
and (\ref{eq:diffEqVar}). Then, set ${\bm \mu}_{k}=\hat{\bm \mu}_{t_1}$
and ${\bm \Sigma}_{k}=\hat{\bm \Sigma}_{t_1}$ and proceed to step 2(b). 
\end{description}
We refer to this modified version of Algorithm 1 as Langevin Incremental Mixture
Importance Sampling (LIMIS). Note that for all practical purposes we can assume that $t_{0}=0$, so the user needs specify only the final time $t_1$. This can be done manually or using the automated approach described in Section \ref{sec:TuningT}. 

LIMIS has several advantageous properties. Firstly, $\hat{\bf \Sigma}_{t_1}$ is guaranteed to be positive definite, even when $\pi({\bf x})$ is not log-concave. This is easily seen by considering discrete-time case, and noticing that the r.h.s. of (\ref{eq:linDisCov}) is positive definite. Secondly, the resulting approximation does not use a
non-parametric estimator, such as Nearest Neighbour, to determine
$\bm \Sigma_{k}$. As will be shown in Section \ref{sec:Examples} this
is especially advantageous in high dimensions. Thirdly, as $t_1$
increases, the mixture components move toward
the nearest mode of $\pi({\bf x})$. This feature has been found to be advantageous by \cite{wlest1991modelling}, who noticed that mixture approximations are typically over-dispersed relative to the target density, and proposed to shrink the mixture components towards the sample mean. As noted by \cite{givens1996local}, West's
method is less appropriate when the target is highly non-Gaussian.
In contrast, we have found that shrinking toward the nearest
mode of $\pi({\bf x})$, by following steepest ascent curves, leads to improved performance even when $\pi({\bf x})$ is far from Gaussian. Finally, the most important property of LIMIS is that it provides an extremely parsimonious parametrization of the mixture importance density. Indeed, the locations and covariances of the mixture components are determined by equations (\ref{eq:diffEqMean}) and (\ref{eq:diffEqVar}). Through these, LIMIS extracts local information about the target, which allows it to limit the number of free parameters that determine the shape of the mixture density to one: the final pseudo-time $t_1$.

\section{Step-size selection} \label{sec:stepSizeSelection}

As explained in Section \ref{sec:linearLangevin}, equations (\ref{eq:diffEqMean})
and (\ref{eq:diffEqVar}) can be used to propagate the mean vector, ${\bm \mu}_t$,
and covariance matrix, ${\bf \Sigma}_t$, of each mixture component between
$t_0$ and $t_1$. In general, the solutions will be approximated using a
numerical integrator, such as a Runge-Kutta scheme. Let $L_{\bm \mu}({\bm \mu},\delta t)$
and $L_{\bf \Sigma}({\bf \Sigma},\delta t)$ be the operators used to
update the moments, that is
\[
\hat{\bm \mu}_{t+\delta t}=L_{\bm \mu}({\bm \mu}_{t},\delta t),\;\;\;\hat{\bf \Sigma}_{t+\delta t}=L_{\bf \Sigma}({\bf \Sigma}_{t},\delta t),
\]
which depend on the numerical scheme used. Here  ${\bm \mu}_{t}$ and
${\bm \Sigma}_{t}$ represent the true solutions of (\ref{eq:diffEqMean}) and
(\ref{eq:diffEqVar}), hence the local truncation errors of the numerical
integrator are
\[
{\bm e}_{{\bm \mu}}={\bm \mu}_{t+\delta t}-\hat{\bm \mu}_{t+\delta t},\;\;\;{\bm e}_{\bf \Sigma}={\bf \Sigma}_{t+\delta t}-\hat{\bf \Sigma}_{t+\delta t},
\]
which are generally $O\{(\delta t)^{\psi}\}$, for $\psi>1$ \citep{suli2003introduction}. While
it is possible to choose $\delta t$ so that numerical estimates of
$|{\bm e}_{{\bm \mu}}|$ and $|{\bm e}_{\bf \Sigma}|$ are below certain thresholds, here
we propose a different approach. In particular, we describe a novel statistically-motivated measure of discretization quality, which we
then use to determine the step-size $\delta t$.

Our proposal consists in quantifying the integration quality in terms
of distance between two local
Gaussian densities: $q({\bf x})=\phi({\bf x}|\hat{\bm \mu}_{t+\delta t},\hat{\bf \Sigma}_{t+\delta t})$
and $q^{*}({\bf x})=\phi({\bf x}|\bm \mu_{t+\delta t},{\bf \Sigma}_{t+\delta t})$. While there are several possible
distance measures that could be adopted, such as the Kullback-Leibler (KL) divergence, we would like a measure that is easily interpretable. For this reason we consider the Population Effective Sample Size (PESS), which we define as
\begin{equation} \label{eq:genPESS}
\text{PESS}\big\{ q({\bf x}),q^{*}({\bf x})\big\}=\underset{n\rightarrow\infty}{\plim} \frac{\text{ESS}^\text{IS}\big\{ q({\bf x}),q^{*}({\bf x})\big\} }{n} =\Bigg [ \int\bigg\{\frac{q({\bf x})}{q^{*}({\bf x})}\bigg\}^{2}q^{*}({\bf x})d{\bf x} \Bigg ]^{-1},
\end{equation} 
where
\begin{equation} \label{eq:ESSIS}
\text{ESS}^\text{IS}\big\{ q({\bf x}),q^{*}({\bf x})\big\} = \bigg\{ \sum_{i=1}^{n} \frac{q({\bf x}_{i})}{q^{*}({\bf x}_{i})}\bigg\}^{2} \bigg / \sum_{i=1}^{n}\bigg\{\frac{q({\bf x}_{i})}{q^{*}({\bf x}_{i})}\bigg\}^{2},
\end{equation}
is the Effective Sample Size (ESS) measure proposed by \cite{kong1994sequential} and ${\bf x}_i \sim q^*({\bf x})$ for $i = 1,\dots,n$. As we show in Appendix \ref{sec:expectedESS}, when both $q({\bf x})$
and $q^{*}({\bf x})$ are Gaussian densities, it is possible to obtain an analytic
expression for the PESS
\begin{eqnarray}
\text{PESS}\big\{ q({\bf x}),q^{*}({\bf x})\big\} & = & \Bigg[\bigg(|2{\bf \Sigma}_{q^*}-{\bf \Sigma}_{q}|\bigg)^{-\frac{1}{2}}|{\bf \Sigma}_{q}|^{-\frac{1}{2}}|{\bf \Sigma}_{q^*}|\nonumber \\
 & \times & \exp\bigg\{(\bm \mu_{q^{*}}-\bm \mu_{q})^{T}(2{\bf \Sigma}_{q^{*}}-{\bf \Sigma}_{q})^{-1}(\bm \mu_{q^{*}}-\bm \mu_{q})\bigg\}\Bigg]^{-1}.\label{eq:PESS}
\end{eqnarray}
This distance measure has the
advantage of having a clear statistical interpretation: it is the limiting value
of the ESS, normalized by the number of samples $n$. Recall that we are solving (\ref{eq:diffEqMean}) and (\ref{eq:diffEqVar}) in order to construct an additional density to be added to the importance mixture. Hence, at each step of the numerical integrator, we are not interested in assessing the accuracy of the approximate solutions $(\hat{\bm \mu}_{t+\delta t}, \hat{\bm \Sigma}_{t+\delta t})$ per se, but we want to quantify how the discretization error perturbs the corresponding density, $q({\bf x})$, away from $q^*({\bf x})$. Therefore, we prefer using (\ref{eq:PESS}), rather the truncation errors ${\bm e}_{{\bm \mu}}$ and ${\bm e}_{\bf \Sigma}$, to determine the steps size. Notice also that $\text{PESS}\big\{ q({\bf x}),q^{*}({\bf x})\big\} \in [0,1]$, as long as $2{\bf \Sigma}_{q^*}-{\bf \Sigma}_{q}$ is positive definite, while $\text{KL}{\{ q({\bf x}),q^{*}({\bf x})\big\}} \geq 0$. Most importantly, by looking at (\ref{eq:genPESS}) it is simple to realize that the chosen criterion is invariant under transformation of $\bf x$, which certainly is not the case for $|\bf{e}_{\bm \mu}|$ and $|\bf{e}_{\bf \Sigma}|$.

Having defined an appropriate distance measure, the step size $\delta t$
can be selected at $t_0$, and kept fixed afterward, or adaptively
at each step. Here we follow the former approach. In particular, if we indicate with $\bm \mu_{t_0}$ and $\bm \Sigma_{t_0}$ the initial moments, then the step-size is selected as follows
\begin{equation}
\delta t^{*}=\bigg[\delta t\;:\;\text{PESS}\big\{\phi({\bf x}|\hat{\bm \mu}_{t_0+\delta t},\hat{\bm \Sigma}_{t_0+\delta t}),\phi({\bf x}|\bm \mu_{t_0+\delta t},\bm \Sigma_{t_0+\delta t})\big\}=\alpha\bigg],\label{eq:critStep-1}
\end{equation}
where the parameter $\alpha\in(0,1)$ is user-defined. Increasing
$\alpha$ reduces the step-size, which leads to more accurate, but computationally more expensive, solutions to (\ref{eq:diffEqMean}) and (\ref{eq:diffEqVar}).
We generally use $\alpha=0.99$ as default value. While the true
moments $({\bm \mu}_{t+\delta t}, {\bm \Sigma}_{t+\delta t})$, needed to compute
(\ref{eq:critStep-1}), are typically unknown, they can be approximated by propagating the moments between $t_0$ and $t_0 + \delta t$ using a smaller step-size, such as $\delta t/10$. Finally, (\ref{eq:critStep-1})
can generally be solved in just a few iterations by a standard one-dimensional
root-finding algorithm, such as Brent's method \citep{brent2013algorithms}.

\section{Examples\label{sec:Examples}}

Here we compare the new LIMIS sampler with NIMIS, IS and MALA. In particular, we use these algorithms to sample a highly non-Gaussian mixture density and the posterior distribution of a Bayesian logistic regression model.

\subsection{Set-up} \label{sec:exSettings}

We compare the performance of the samplers using several criteria. While some of these, such as Root Mean Squared Errors (RMSEs), are well known, others are less well known and so specified here. In Section \ref{sec:banExample} we evaluate the methods using the marginal accuracy measure of \cite{faes2012variational}, that is 
\[
\text{MA}=1-\frac{1}{2}\int_{-\infty}^{+\infty}|\pi({\bf x})-\hat{\pi}({\bf x})|d{\bf x},
\]
where $\text{MA}=1$ if $\pi({\bf x})$ and $\hat{\pi}({\bf x})$ are identical and $\text{MA}=0$ if the two densities do not overlap anywhere. When weighted samples ${\bf z}_1,\dots,{\bf z}_n$, drawn from $q({\bf z})$, are available, $\pi({\bf x})$ is estimated by
\[
\hat{\pi}({\bf x})=\frac{1}{hn}\sum_{i=1}^{n}\kappa_{h}({\bf x}|{\bf z}_i)w_{i}\approx \int_{-\infty}^{+\infty}\kappa_{h}({\bf x}|{\bf z})\frac{\pi({\bf z})}{q({\bf z})}q({\bf z})d{\bf z},
\]
with $\kappa_{h}({\bf x}|{\bf z})$ being a kernel density, with bandwidth $h$. An additional criterion is efficiency (EF), by which we indicate the ratio of ESS to total number of samples $n$. For LIMIS, NIMIS and IS we use formula (\ref{eq:ESSIS}) to compute the ESS, 
while for MALA we use 
$$
\text{ESS}^{\text{MC}} = \frac{n}{1+2\sum_{t=1}^\infty \rho_t},
$$
where $\rho_t$ is the autocorrelation of the chain at lag $t$. Notice that under both definitions $\text{ESS}\in[1,n]$, so $\text{EF}\in[0,1]$.

We report RMSEs of estimated marginal means, variances and normalizing constant of the target ($\int\pi({\bf x})d{\bf x}$). While estimating the normalizing constant is straightforward when importance samples are available, much more care is required when using MCMC methods. Hence, we do not estimate this quantity when applying MALA.  

In terms of algorithmic parameters, for LIMIS and NIMIS we use $\nu=3$, which is the smallest integer value of $\nu$ such that the variance of a Student's t random variable is finite, and we follow \cite{raftery2010estimating} who 
suggest the default values $n_{0}=1000d$, $b=100d$. We use an equal number ($n_0+kb$) of samples or iterations for IS and MALA. The step size of LIMIS is determined as explained in Section \ref{sec:stepSizeSelection}. The only LIMIS parameter that we chose manually is the final pseudo-time $t_1$. However, we discuss how it can be selected in an automated fashion in Section \ref{sec:TuningT}. When applying MALA we discard the first tenth of each MCMC chain as the burn-in period, and we select the step size so as to approximately achieve the optimal 0.574 acceptance rate derived by \cite{roberts2001optimal}.   The remaining settings will be detailed in Sections \ref{sec:banExample} and \ref{sec:logRegr}.

All the examples are implemented in the \verb|Julia| language \citep{bezanson2012julia}. We developed our own implementation of LIMIS and NIMIS, while we use the MALA algorithm offered by the \verb|Klara| \verb|Julia| package.

\subsection{Mixture of warped Gaussians} \label{sec:banExample}

\begin{figure}
\centering{}\includegraphics[scale=0.35]{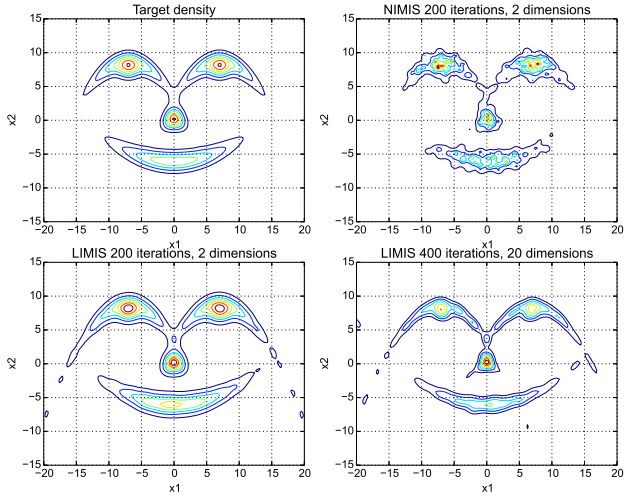}
\caption{Two-dimensional slice of the target and of the importance densities, obtained using LIMIS and NIMIS.}\label{fig:clownPlot}
\end{figure}

As a first example we consider a mixture target density 
\[
\pi({\bf x})=\sum_{i=1}^{r}w_{i}p_{i}({\bf x}),\;\;\;\sum_{i=1}^{r}w_{i}=1,
\]
where each of the $r$ mixture components is a shifted version of the banana-shaped
density described in \cite{haario2001adaptive}. In particular, let
${\bf y}\sim N(0,{\bm \Sigma}_{a})$, where ${\bm \Sigma}_{a}=\text{diag}(a^{2},1,\dots,1)$,
and consider the following transformed random variables 
\[
x_{1}=y_{1}+s_{1},\;\;\;x_{2}=y_{2}-b(y_{1}^{2}-a^{2})+s_{2},\;\;\;x_{i}=y_{i},\;\;\;\text{for}\;\;\;i=3,\dots,d,
\]
where $a$, $b$, $s_{1}$ and $s_{2}$ are constants. Given that
the determinant of the Jacobian of this transformation is $1$, the
density of ${\bm x}$ is simply
\[
p({\bf x})=\phi[x_{1}-s_{1},x_{2}+b\{(x_{1}-s_{1})^{2}-a^{2}\}-s_{2},x_{3}\dots,x_{d}|{\bm 0},{\bm \Sigma}_{a}],
\]
where $\phi({\bf x}|{\bm \mu},{\bm \Sigma})$ is the p.d.f. of a multivariate normal distribution and ${\bm 0}$ is a $d$-dimensional vector of zeros. We consider a mixture of $r=6$ such densities, each with different values for parameters $a,$ $b$, $s_{1}$ and $s_{2}$. These are reported in
Appendix \ref{sec:BanDeriv}, together with formulas for the gradient and Hessian of $\log\pi({\bf x})$. A slice of the target density across the first two dimensions is shown in top-left plot of Figure \ref{fig:clownPlot}.

We sample $\pi({\bf x})$ using LIMIS, NIMIS, IS and MALA. In particular, we consider three scenarios where $d$ is respectively equal to $5$, $20$ and $80$. For LIMIS and NIMIS we use $k=200$ iterations, and for the former method we let $t_1$ grow with $d$, by setting it to 1, 3 and 5. To initialize LIMIS and NIMIS we use a diffuse $\text{mvt}({\bf x}|{\bf 0},100{\bf I}, 3)$ prior distribution. To perform IS we use a weighted mixture of four multivariate Student's t distributions each centered at one of the modes of the target, ${\bf x}_1^*, \dots, {\bf x}_4^*$, and with covariances equal to $-2\{\nabla^2\log \pi({\bf x})\}|_{{\bf x} = {\bf x}_i^*}^{-1}$, for $i=1,\dots,4$. The weights are reported in Appendix \ref{sec:BanDeriv}. See Section \ref{sec:exSettings} for additional information about the simulation setting.

Table \ref{tab:mseBanana} reports the results obtained using 16 independent runs of each sampler. We do not report the efficiency of MALA, because the ESS was extremely low in all runs, even when the algorithm was mixing properly. In fact, sample autocorrelations are high when the sampler explores the same mode for several hundred iterations before jumping to another mode, which results in extremely low autocorrelation-adjusted ESS estimates. Hence, we consider the resulting efficiency estimates to be quite misleading.

In the five-dimensional scenario NIMIS closely follows LIMIS, which is the best performer on most criteria. However, the performance of NIMIS degrades rapidly as the dimensionality increases, to the point that it failed entirely in 80 dimensions. IS performed quite poorly in all scenarios, especially in terms of efficiency. LIMIS seems to be scaling well with $d$ on most criteria, the estimated marginal variances, $\sum_{i=3}^{d}\text{Var}(x_{i})$, being an exception. Here MALA achieves a lower MSE than LIMIS when $d=20$, and the gap increases when $d=80$. However, in 80 dimensions, LIMIS is still more accurate than MALA in terms of marginal accuracies and of RMSE for $\sum_{i=3}^{d}\mathbb{E}(x_{i})$. 

\begin{table}
\centering
\begin{tabular}{|c|c|c|c|c|c|}
\hline 
5d & LIMIS & NIMIS & IS & MALA & ord. mag. \tabularnewline
\hline 
\hline 
$\text{MA}(x_{1})$ & \textbf{0.991} & {0.989} & 0.965 & 0.954 & 1 \tabularnewline
\hline 
$\text{MA}(x_{2})$ & \textbf{0.990} & {0.988} & 0.967 & 0.929 & 1 \tabularnewline
\hline 
$\sum_{i=3}^{d}\mathbb{E}(x_{i})$ & {0.53}(0.99)  & \textbf{0.62}(0.98)  & 1.74(0.98) & 0.84(0.91)  & $10^{-2}$ \tabularnewline
\hline 
$\sum_{i=3}^{d}\text{Var}(x_{i})$ & \textbf{9.11}(0.97) & {15.96}(0.42) & 33.12(0.97) & 10.91(0.85) & $10^{-3}$  \tabularnewline
\hline 
$\int \pi({\bf x})d{\bf x}$ & \textbf{2.30}(0.35) & {5.80}(0.17) & 13.26(0.99) & - & $10^{-3}$\tabularnewline
\hline 
Efficiency & \textbf{0.69}(0.68)  & {0.52}(0.51) & 0.05(0.02) & - & 1 \tabularnewline
\hline
\hline
20d &   & & & & \tabularnewline
\hline 
$\text{MA}(x_{1})$ & \textbf{0.994} & 0.846 & 0.942 & {0.970} & 1 \tabularnewline
\hline 
$\text{MA}(x_{2})$ & \textbf{0.993} & 0.844 & 0.943  & {0.966}  & 1 \tabularnewline
\hline 
$\sum_{i=3}^{d}\mathbb{E}(x_{i})$ & \textbf{0.97}(0.99) & 8.58(0.73) & 11.56(0.94) & {1.54}(0.98) & $10^{-2}$ 
\tabularnewline
\hline 
$\sum_{i=3}^{d}\text{Var}(x_{i})$ & {47.73}(0.29) & 4429(0.01) & 771(0.94) & \textbf{20.67}(0.85) & $10^{-3}$ \tabularnewline
\hline 
$\int \pi({\bf x})d{\bf x}$ & \textbf{2.45}(0.25) & 306.4(0.01) & {57.56}(0.98) & - & $10^{-3}$\tabularnewline
\hline 
Efficiency & \textbf{0.416}(0.409) & 0.005(0.003) & {0.008}(0.001) & - & 1\tabularnewline
\hline 
\hline
80d &  & & & &  \tabularnewline
\hline 
$\text{MA}(x_{1})$ & \textbf{0.995} & - & 0.945 & {0.982}  & 1 \tabularnewline
\hline 
$\text{MA}(x_{2})$ & \textbf{0.995} & - & 0.947 & {0.980}  & 1\tabularnewline
\hline 
$\sum_{i=3}^{d}\mathbb{E}(x_{i})$ & \textbf{1.13}(0.88) & - & 32.14(0.86) & {2.8}(0.96) & $10^{-2}$ \tabularnewline
\hline 
$\sum_{i=3}^{d}\text{Var}(x_{i})$ & {113.2}(0.35)  & -  & 3029(0.37) & \textbf{27.2}(0.99) & $10^{-3}$ \tabularnewline
\hline 
$\int \pi({\bf x})d{\bf x}$ & \textbf{1.6}(0.37) & -  & {41.1}(0.50) & - & $10^{-3}$ \tabularnewline
\hline 
Efficiency & \textbf{0.22}(0.21) & - & {0.002}(0.0001) & - & 1\tabularnewline
\hline 
\end{tabular}\caption{For each dimension: a) the first two rows report marginal accuracies along the first two dimensions; b) the following three rows contain RMSEs and, between brackets, the ratio between squared bias and MSE; c) the last row reports mean efficiencies and, between brackets, the lowest efficiencies on the 16 runs. For each row, the order of magnitute of the marginal accuracies, RMSEs and efficiencies is reported in the last column. }
\label{tab:mseBanana}
\end{table}

\subsection{Logistic Regression}\label{sec:logRegr}

Here we consider a Bayesian logistic regression problem. Assume we have $n$ i.i.d samples of binary labels ${\bf y} \in\{0,1\}^{n}$ and a corresponding $n\times d$ matrix of covariates $\bf X$.  Under a logistic regression model
$$
\text{Prob}(y_i=1|{\bf X}, \bm \theta)=\frac{e^{{\bf X}_{i:}^T\bm \theta}}{1+e^{{\bf X}_{i:}^T\bm \theta}}, \;\;\; \text{for} \; i=1,\dots,n,
$$ 
where $\bm \theta$ is a vector of model coefficients and ${\bf X}_{i:}^T$ is the $i$-th row of $\bf X$. If $X_{j1} = 1$, for $j = 1, \dots, n$, then $\theta_1$ represents the intercept. If we use a flat prior on $\theta_1$ and a Gaussian prior on $\{ {\bm \theta}_{2}, \dots, {\bm \theta}_{d}\}$, with mean zero and covariance ${\bf I}\lambda^{-1}$, where $\bf I$ is a $d-1$ dimensional identity matrix and $\lambda>0$, the posterior log-density of the parameters is
\[
\log \pi(\bm \theta) \propto {\bf y}^{T}{\bf X}\bm \theta-\sum_{i=1}^{n}\log(1+e^{{\bf X}_{i:}^T\bm \theta})-\frac{\lambda}{2} \sum_{j=2}^d \theta_2^2.
\]
 Formulas for the gradient and Hessian of $\log \pi(\bm \theta)$ are provided in Appendix \ref{app:logRegDetails}.

\begin{table}
\centering
\begin{tabular}{|c|c|c|c|c|c|}
\hline 
 & LIMIS & NIMIS & IS & MALA & scale \tabularnewline
\hline 
\hline 
$\mathbb{E}(\theta_j)$ & \textbf{4.7}(0.95) & 22.5(0.87) & 5.9(0.94) & 11.1(0.95) & $10^{-4}$ \tabularnewline
\hline 
$\text{Var}(\theta_j)^{1/2}$ & \textbf{3.3}(0.94) & 12.9(0.91) & 4.1(0.94) & 5.9(0.92) & $10^{-4}$ \tabularnewline
\hline 
$\int \pi(\bm \theta)d\bm \theta$ & \textbf{2.3}(0.97) & 50.0(0.05) & 3.3(0.89) & - & $10^{-3}$\tabularnewline
\hline 
Efficiency & \textbf{0.18}(0.17) & 0.01(0.01) & 0.11(0.11) & 0.03(0.026) & $1$\tabularnewline
\hline 
\end{tabular}\caption{First three rows: RMSE and, between brackets, the ratio between squared bias and MSE for each estimate and method. Last row: mean and, between brackets, minimal efficiency on the 16 runs. Last column: order of magnitude of the RMSEs.}
\label{tab:MSElogReg}
\end{table}

To verify how LIMIS performs on this model, we consider the Sonar dataset, which is freely available within the UCI repository \citep{Lichman:2013}. The dataset was originally considered by \cite{gorman1988analysis}, who used it to train a neural network to discriminate sonar signals bounced off a mine from those bounced of a rock. It includes $n=208$ observations, where the response variable indicates whether the object is a mine ($y=1$) or a rock ($y=0$). Each covariate vectors contains $d=60$ numbers ranging between 0 and 1, which represent the signal's energy within a specific frequency interval, integrated over time. See \cite{gorman1988analysis} for more details on the dataset. 

We aim at sampling $\pi(\bm \theta)$ using LIMIS, NIMIS, IS and MALA, for fixed $\lambda$. After standardizing the features $\bf X$, we select $\lambda \approx 28$, by $k$-fold cross-validation. For LIMIS and NIMIS we use $k=100$ iterations, and for MALA we discard the first $10\%$ of each chain as burn-in period. As importance distribution for IS we use a multivariate Student's t distribution with 3 degrees of freedom, centred at the posterior mode, $\bm \theta^*$, and with covariance matrix equal to $-2\{\nabla^2\log \pi(\bm \theta)\}|_{\bm \theta = \bm \theta^*}^{-1}$. We use the same density to initialize LIMIS and NIMIS. The remaining settings are as described in Section \ref{sec:exSettings}.

Table \ref{tab:MSElogReg} summarizes the results of 16 independent estimation runs. The first three rows report the RMSEs of the estimated marginal posterior means and variances, averaged over the 61 dimensions, and of the estimated normalizing constant or marginal likelihood. Notice that LIMIS is the best performing method here, followed by IS. IS does well because the chosen value of $\lambda$ results in an approximately Gaussian posterior. Given the results obtained in Section \ref{sec:banExample}, we expect that lowering $\lambda$ would make the target less Gaussian, thus narrowing the performance gap between IS and MALA. As expected, NIMIS is the worst performing method, due to the high dimensionality of the problem. 

\section{Computational considerations} \label{sec:computCons}

In the previous examples we have not reported the computing times of the different methods, because these are highly implementation dependent. However, here we make some general considerations about computational efficiency, which are less dependent on the software implementation.

Let $n_{k}=n_{0}+kb$ be the total number of samples obtained using LIMIS and NIMIS, where $k$ is the number of iterations, $n_0$ is the number of samples from the prior and $b$ is the number of samples simulated at each iteration. Assume that the total number of samples obtained with MALA and IS is also $n_k$. An important factor in determining the attractiveness of each method is the cost of evaluating $\log \pi(\bf x)$ and its derivatives. MALA requires $n_k$ evaluations of $\nabla \log \pi(\bf x)$. LIMIS evaluates gradient and Hessian several times when constructing the $k$ mixture densities. The factor multiplying $k$ depends on the number of steps used in the Langevin linearization of Section \ref{sec:linearLangevin}. For instance, using the default $\alpha$ proposed in Section \ref{sec:stepSizeSelection} to determine the step size, $\delta t$, the linearization requires on average 50 integration steps in the twenty-dimensional warped Gaussian mixture example. In that scenario we used $k=200$, $b=100d$ and $n_0=1000d$, hence a whole LIMIS run requires around $kd=10^4$ evaluations of gradient and Hessian, which should be compared with the $n_0+kb=42\times10^4$ gradient evaluations required by MALA. The Hessian of this example is highly sparse but, for a typical model, computing it should be $O(d)$ times more expensive than evaluating the gradient. Hence, if the Hessian of this example was dense, the total cost of computing the derivatives under LIMIS and MALA would roughly match. However, notice that LIMIS outputs a mixture density which can be used to do further importance sampling, and this does not require any additional derivative evaluation.

\begin{figure}
\centering{}\includegraphics[scale=0.35]{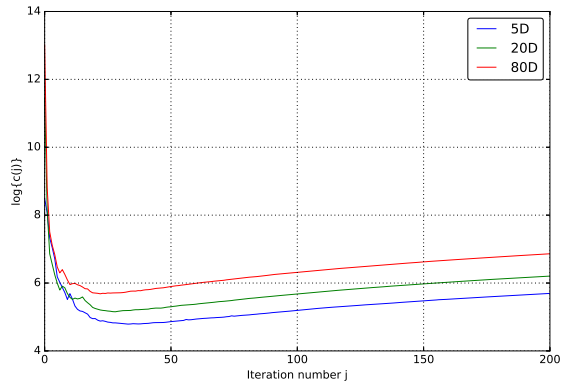}
\caption{Log-cost per sample, averaged over 16 runs, under the three scenarios considered in Section \ref{sec:banExample}.}\label{fig:cost}
\end{figure}

A second factor is the cost of evaluating the importance density.
At the $j$-th iteration of LIMIS or NIMIS, where $j\in\{1,\dots,k\}$, the cost of single evaluation is $O(jd^2)$. While in the examples we ran these algorithms until a fixed $k$ was reached, it might be preferable to stop when the increased cost of evaluating the importance mixture is not more than offset by gains in efficiency. In particular, let $c_\pi$ and $c_q$ be, respectively, the cost of evaluating the density of the target or of a single mixture component. Then the cost of an independent sample is approximately
\begin{equation} \label{eq:stopCrit}
c(j) = \frac{c_\pi + j c_q}{\text{EF}(j)},\;\;\; \text{for} \;\; j=1,2,\dots,
\end{equation}
where $\text{EF}(j)$ is the efficiency of an importance mixture with $j$ components. Figure \ref{fig:cost} shows the behaviour of $c(j)$ when running LIMIS on the mixture example. For more complex examples accurate time estimates would be required, but here the target is a mixture of six warped Gaussian densities, hence we assumed $c_\pi \approx 6c_q$. The plot suggests that the computational budget could be used more efficiently by stopping LIMIS around the 25th iteration, and using the resulting mixture importance density to obtain more samples. 

In the previous examples we have seen that, from the point of view of statistical efficiency, the performance of NIMIS is very unsatisfactory in high dimensions. IS scales better, as long as a good approximation to the target is available, as in the logistic example. IS is also computationally cheap, because it does not require derivative information. However, in high-dimensional non-Gaussian scenarios a good off-the-shelf importance distribution is, in most cases, not readily available. In these cases, using LIMIS to construct an efficient importance distribution might be advantageous. In fact, the output mixture density could then be used to obtain more importance samples, which would not require any further evaluations of the target's gradient and Hessian. Obviously, one has to be careful not grow the size of the importance mixture to the point that the increase in the cost of evaluating this density is not justified by the resulting statistical efficiency gains. This could be avoided by stopping LIMIS when a criterion such as (\ref{eq:stopCrit}) is approximately minimized.

\section{Tuning the final pseudo-time $t_1$} \label{sec:TuningT}

In the examples presented in Section \ref{sec:Examples} we selected the final pseudo-time $t_1$ manually. In general we start from the default $t_1=1$ and check whether perturbing $t_1$ drastically improves a performance measure, such as EF. In the logistic regression example the performance did not seem to depend much on $t_1$, hence we used its default value. In the mixture density example we increased $t_1$ with the dimensionality, $d$, of the target. Increasing $t_1$ inflates the covariance of the importance mixture components and it shinks their locations towards the closest mode of the target. Given that distances increase with $d$, this is a desirable behaviour. An alternative approach would have been to increase the number of LIMIS iterations with $d$, while keeping $t_1$ constant. 

In this section we show how an initial choice of $t_1$ can be improved, using a fully automated procedure. Assume that the results
of a preliminary LIMIS run are available, which include a weighted sample ${\bf x}_1,\dots,{\bf x}_{n_k}$, where $n_{k}=n_{0}+kb$
and $k$ is the number of LIMIS iterations. Indicate with $\tilde{\bf x}_1,\dots, \tilde{\bf x}_k$ the samples that achieved the highest weight in one of the iterations, and hence resulted in the addition of a mixture component. The mean vectors, $\bm \mu^{1}_{t_1},\dots,\bm \mu^{k}_{t_1}$, and covariance matrices, $\bm \Sigma^{1}_{t_1}, \dots, \bm \Sigma^{k}_{t_1}$, of these components are constructed using the Langevin linearization methods of Section \ref{sec:linearLangevin}, and thus depend on the pseudo-time $t_1$. Here we denote $q({\bf x}|t_1)$ as the resulting mixture density. In this section we aim at selecting $t_1$ so that $q({\bf x}|t_1)$ is optimal, in a sense to be clarified shortly.

Suppose that we wish to estimate
\[
I=\mathbb{E}\big\{ h({\bf x})\big\}=\int h({\bf x})\frac{\pi({\bf x})}{c}d{\bf x} = \int h({\bf x})\tilde{\pi}({\bf x})d{\bf x} ,
\]
where $c = \int \pi({\bf x})d{\bf x}$ and $h({\bf x})$ is an $\mathbb{R}^d\rightarrow\mathbb{R}$ function.  If only the un-normalized target, $\pi({\bf x}) = c\,\tilde{\pi}({\bf x})$, can be evaluated, then $I$ can be estimated by self-normalized importance sampling, that is
\[
\hat{I}=\frac{\sum_{j=1}^{m}h({\bf x}_{j})w_j}{\sum_{j=1}^{m}w_j}, \;\;\;\text{where}\;\;\;w_j = \frac{\pi({\bf x}_{j})}{q({\bf x}_{j}|t_1)}\;\;\;\text{and}\;\;\;{\bf x}_{j}\sim q({\bf x}_{j}|t_1),
\]
for $j=1,\dots,m$. The asymptotic variance of $\hat{I}$ is proportional to
\begin{equation}
v(t_1) = \frac{\int\frac{\pi({\bf x})^2}{q({\bf x}|t_1)^2}\big\{h({\bf x})-I\big\}^2q({\bf x}|t_1)d{\bf x}}{\big\{\int \frac{\pi({\bf x})}{q({\bf x}|t_1)} q({\bf x}|t_1)d{\bf x}\big\}^2}, \label{eq:optimCritUNN}
\end{equation}
hence, ideally, we would like to determine the value, $t_1^*$, that minimizes
(\ref{eq:optimCritUNN}).
In order to approximately achieve this, we need a reasonably
cheap estimator of $v(t_1)$. Let $q({\bf x}|t_1^{I})$ be the mixture importance density in the final iteration of the pilot LIMIS run. Then (\ref{eq:optimCritUNN}) can be estimated by
\begin{equation} \label{eq:tuneTformUNN}
\hat{v}(t_1) = \frac{\frac{1}{n_k}\sum_{i=1}^{n_{k}}\frac{\pi({\bf x}_i)^2}{q({\bf x}_i|t_1)^2}\big\{h({\bf x}_i)-\hat{I}\big\}^2\frac{q({\bf x}_i|t_1)}{q({\bf x}_i|t_1^I)}}{\big\{\frac{1}{n_k}\sum_{i=1}^{n_{k}} \frac{\pi({\bf x}_i)}{q({\bf x}_i|t_1)} \frac{q({\bf x}_i|t_1)}{q({\bf x}_i|t_1^I)}\big\}^2} = \frac{1}{\hat{c}^2n_k}\sum_{i=1}^{n_{k}}\frac{\pi({\bf x}_i)}{q({\bf x}_i|t_1)}\big\{h({\bf x}_i)-\hat{I}\big\}^2w_i,
\end{equation}
where
\begin{equation} \label{eq:forLaterUse}
\hat{I}=\frac{1}{\hat{c}\,n_k} \sum_{i=1}^{n_{k}}h({\bf x}_{i})w_i, \;\;\; \hat{c} = \frac{1}{n_k} \sum_{i=1}^{n_{k}} w_i,  \;\;\; w_i = \frac{\pi({\bf x}_{i})}{q({\bf x}_{i}|t_1^I)},\;\;\;\text{and}\;\;\;{\bf x}_{i}\sim q({\bf x}_{i}|t_1^I),
\end{equation}
for $i=1,\dots,n_{k}$. Here ${\bf x}_{i}$, $h({\bf x}_{i})$, $\pi({\bf x}_{i})$, $q({\bf x}_{i}|t_{1}^I)$, $w_i$, $\hat{I}$ and $\hat{c}$ have already been simulated/computed and stored during the preliminary run. Hence $\hat{v}(t_1)$ is a deterministic function, which can be minimized using a one-dimensional optimizer, where only $q({\bf x}_{i}|t_1)$, for $i=1,\dots,n,$ needs to be recomputed as the optimizer explores different values of $t_1$. If the normalized target, $\tilde{\pi}({\bf x})$, can be computed directly and $I$ is estimated using 
\[
\tilde{I}=\frac{1}{m}\sum_{j=1}^{m}h({\bf x}_{j})\frac{\tilde{\pi}({\bf x}_{j})}{q({\bf x}_{j}|t_1)}, \;\;\;\text{where}\;\;\;{\bf x}_{j}\sim q({\bf x}_{j}|t_1),\;\;\;\text{for}\;\;\; j=1,\dots,m,
\]
then the finite-sample variance of $\tilde{I}$ is proportional to 
\begin{equation}
\tilde{v}(t_1)=\int\frac{h({\bf x})^{2}\tilde{\pi}({\bf x})^{2}}{q({\bf x}|t_1)}d{\bf x}, \label{eq:optimCrit}
\end{equation}
which can be estimated by
\begin{equation} \label{eq:tuneTform}
\hat{\tilde{v}}(t_1)=\frac{1}{n_k}\sum_{i=1}^{n_{k}}\frac{h({\bf x}_{i})^2\tilde{\pi}({\bf x}_{i})}{q({\bf x}_{i}|t_1)}w_{i},\;\;\;\text{where}\;\;\;w_i=\frac{\tilde{\pi}({\bf x}_{i})}{q({\bf x}|t_{1}^I)}\;\;\;\text{and}\;\;\;{\bf x}_{i}\sim q({\bf x}|t_{1}^I),
\end{equation}
for $i=1,\dots,n_{k}$. Also in this case only $q({\bf x}_i|t_{1})$ needs to be recomputed at $t_1$ varies.

Notice that, if we set $h({\bf x})=1$ in (\ref{eq:optimCrit}), minimizing $\tilde{v}(t_1)$ is equivalent to maximizing $\text{PESS}\{\tilde{\pi}({\bf x}), q({\bf x}|t_1)\}$ (\ref{eq:genPESS}). This choice is useful when the practitioner is not interested in minimizing the variance under any particular integrand $h({\bf x})$, but wants to obtain an importance density that is adapted to the target. However, in the self-normalized case, setting $h({\bf x})=1$ leads to $v(t_1)=\hat{v}(t_1)=0$ for any $t_1$, because this estimator is exact for constant $h({\bf x})$. Hence, if the normalizing constant is unknown and no specific $h({\bf x})$ is particularly relevant, then $v(t_1)$ might not be the best criterion to use. An alternative is to consider the Kullback-Leibler (KL) divergence between $\tilde{\pi}({\bf x})$ and $q({\bf x}|t_1)$, that is
\begin{equation} \label{eq:kldist}
\text{KL}(t_1) = \int \log\bigg\{\frac{\tilde{\pi}({\bf x})}{q({\bf x}|t_1)}\bigg\} \tilde{\pi}({\bf x}) d{\bf x} \propto - \int \log\big\{q({\bf x}|t_1)\big\} \frac{\pi({\bf x})}{c} d{\bf x},
\end{equation}
as done, in a related context, by \cite{cappe2008adaptive}. The r.h.s. of (\ref{eq:kldist}), which we indicate with $g(t_1)$, can be estimated by
\begin{equation} \label{eq:klestim}
\hat{g}(t_1)=-\frac{1}{\hat{c}\,n_k}\sum_{i=1}^{n_{k}}\log\big\{q({\bf x}_i|t_1)\big\}w_{i},\;\;\;\text{where}\;\;\;{\bf x}_{i}\sim q({\bf x}|t_{1}^I),\;\;\; \text{for}\;\;\; i=1,\dots,n_{k},
\end{equation}
with $\hat{c}$ and the $w_i$s being defined as in (\ref{eq:forLaterUse}).

\begin{figure}
\centering{}\includegraphics[scale=0.35]{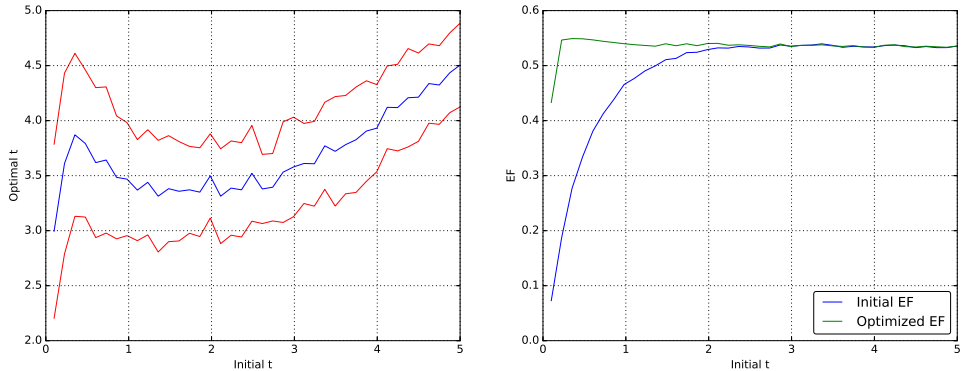}
\caption{Left: mean ($\pm \sigma$) optimal pseudo-time $t_1^*$, as a function of initialization $t_1^I$. Right: mean efficiency (EF) when the mixture derived using the initial, $t_1^I$, or the optimized, $t_1^*$, pseudo-time is used for importance sampling.}\label{fig:tuneT}
\end{figure}

To provide a simple illustration, we consider again the mixture target density of Section \ref{sec:banExample}. In particular, we set $d=5$ and we run LIMIS for $k=50$ iterations, using a grid of initial values for $t_1^I$. We then estimate the optimal value of $t_1$ by minimizing $\hat{\tilde{v}}(t_1)$ with $h({\bf x})=1$. To reduce the computational effort, we compute (\ref{eq:tuneTform}) using a sub-sample of size $n_k/10$, drawn multinomially from the $n_k$ available samples. The left plot in Figure \ref{fig:tuneT} shows, for each value of $t_1^I$, the estimated $t_1^*$, averaged over 60 runs. After estimating $t_1^*$ for each $t_1^I$, we use each of the resulting mixture densities within an importance sampler, and we evaluate its efficiency. The average efficiencies of $q({\bf x}|t_1^*)$ and $q({\bf x}|t_1^I)$ are compared in the right plot of Figure \ref{fig:tuneT}. Optimizing over $t_1$ brings about drastic improvements in efficiency, if $t_1^I$ is set too low. This is to be expected, because for low $t_1$ the importance mixture density is composed of widely spaced and narrow modes, which leads to highly variable weights.

\section{Conclusions} \label{sec:conclusion}

The LIMIS algorithm provides a simple but flexible iterative framework for concurrently constructing a mixture importance density and performing importance sampling using such a density. By exploiting the local information about the target density, LIMIS scales well with the dimensionality of the sampling space, especially if compared with the original NIMIS algorithm. The examples show that the performance of LIMIS compares favorably with that of a state-of-the-art MCMC sampler such as MALA, under either a nearly Gaussian (Section \ref{sec:logRegr}) and a multimodal (Section \ref{sec:banExample}) target. 

In Section \ref{sec:TuningT} we showed how the final pseudo-time, $t_1$, can be selected by minimizing a function-specific variance estimate. This requires post-processing the results of a preliminary LIMIS run. In addition, notice that in this work we assumed that the components of the importance mixture are equally weighted. Perhaps the most promising direction for future research would be adaptively selecting $t_1$ and, possibly, the mixture weights at each iteration of the sampler. \cite{cappe2008adaptive} select weights and parameters of a mixture of Gaussian or multivariate Student's t densities, by adaptively minimizing an entropy criterion. We think that their approach could be adjusted to fit our context. We expect that the resulting adaptive algorithm would benefit greatly from the fact that the locations and covariance matrices of LIMIS mixture components are entirely controlled by the target's shape and by $t_1$, which would drastically reduce the number of parameters that need to be optimized during the adaptation step.

\section{Aknowledgments} \label{sec:conclusion} 
The authors would like to thank Samuel Livingstone for useful comments on an early draft of this paper. Most of this work was undertaken at the University of Liverpool, where M.F. was a post-doctoral research associate, and it was funded by Dstl through projects WSTC0058 and CDE36610.


\bibliographystyle{chicago}
\bibliography{biblio.bib}


\begin{appendices}

\section{Derivation of the Population Effective Sample Size} \label{sec:expectedESS}

Consider a Gaussian importance density, with mean ${\bm \mu}_{1}$ and covariance
${\bm \Sigma}_{1}$, and a Gaussian target, with mean ${\bm \mu}_{2}$ and covariance
${\bm \Sigma}_{2}$. Then the PESS is defined by
\[
\text{PESS}\{\text{\ensuremath{\phi}}({\bf x}|{\bm \mu}_{2},{\bm \Sigma}_{2}), \text{\ensuremath{\phi}}({\bf x}|{\bm \mu}_{1},{\bm \Sigma}_{1}) \}=\bigg\{\int\bigg[\frac{\text{\ensuremath{\phi}}({\bf x}|{\bm \mu}_{2},{\bm \Sigma}_{2})}{\text{\ensuremath{\phi}}({\bf x}|{\bm \mu}_{1},{\bm \Sigma}_{1})}\bigg]^{2}\text{\ensuremath{\phi}}({\bf x}|{\bm \mu}_{1},{\bm \Sigma}_{1})\,d{\bf x}\bigg\}^{-1}=\mathbb{E}(w^{2})^{-1}.
\]
Simple manipulations lead to
\begin{eqnarray*}
\mathbb{E}(w^{2}) & = & (2\pi)^{-\frac{d}{2}}\frac{|{\bm \Sigma}_{1}|^{\frac{1}{2}}}{|{\bm \Sigma}_{2}|}\int\exp\bigg\{-({\bf x}-{\bm \mu}_{2})^{T}{\bm \Sigma}_{2}^{-1}({\bf x}-{\bm \mu}_{2})+\frac{1}{2}({\bf x}-{\bm \mu}_{1})^{T}{\bm \Sigma}_{1}^{-1}({\bf x}-{\bm \mu}_{1})\bigg\} d{\bf x} \nonumber \\
& = & \bigg(2{}^{d}|-{\bf A}|\bigg)^{-\frac{1}{2}}\frac{|{\bm \Sigma}_{1}|^{\frac{1}{2}}}{|{\bm \Sigma}_{2}|}e^{{\bf c}-\frac{1}{4}{\bf b}^{T}{\bf A}^{-1}{\bf b}},
\end{eqnarray*}
where 
$$
{\bf A} = \frac{1}{2}{\bm \Sigma}_{1}^{-1}-{\bm \Sigma}_{2}^{-1}, \;\;\;\; {\bf b} = 2{\bm \Sigma}_{2}^{-1}{\bm \mu}_{2}-{\bm \Sigma}_{1}^{-1}{\bm \mu}_{1} \;\;\; \text{and} \;\;\; {\bf c} = \frac{1}{2}{\bm \mu}_{1}^{T}{\bm \Sigma}_{1}^{-1}{\bm \mu}_{1}-{\bm \mu}_{2}^{T}{\bm \Sigma}_{2}^{-1}{\bm \mu}_{2}.
$$
The exponent can be simplified, in fact the properties $\text{Tr}(\bf X + \bf Y) = \text{Tr}(\bf X) + \text{Tr}(\bf Y)$ and $\text{Tr}({\bf XYZ})=\text{Tr}({\bf ZXY}) = \text{Tr}({\bf YZX})$ lead to
\begin{eqnarray*}
{\bf c}-\frac{1}{4}{\bf b}^{T}{\bf A}^{-1}{\bf b} & = & \text{Tr}\Big ( {\bf c}-\frac{1}{4}{\bf b}^{T}{\bf A}^{-1}{\bf b} \Big ) \\
& = & \text{Tr}\Bigg\{\bigg(\frac{1}{2}{\bm \Sigma}_{1}^{-1}-{\bm \Sigma}_{2}^{-1}\bigg)^{-1}\bigg(-\frac{1}{2}{\bm \Sigma}_{1}^{-1}{\bm \mu}_{2}{\bm \mu}_{2}^{T}{\bm \Sigma}_{2}^{-1}-\frac{1}{2}{\bm \Sigma}_{2}^{-1}{\bm \mu}_{1}{\bm \mu}_{1}^{T}{\bm \Sigma}_{1}^{-1}\\
& + & \frac{1}{2}{\bm \Sigma}_{1}^{-1}{\bm \mu}_{1}{\bm \mu}_{2}^{T}{\bm \Sigma}_{2}^{-1}+\frac{1}{2}{\bm \Sigma}_{2}^{-1}{\bm \mu}_{2}{\bm \mu}_{1}^{T}{\bm \Sigma}_{1}^{-1}\bigg)\Bigg\}.
\end{eqnarray*}
Then we can use the property
\begin{equation} \label{eq:matrixInverse}
\big({\bf X}^{-1}+{\bf Y}^{-1}\big)^{-1}={\bf X}\big({\bf X}+{\bf Y}\big)^{-1}{\bf Y}={\bf Y}\big({\bf X}+{\bf Y}\big)^{-1}{\bf X},
\end{equation}
to obtain
\begin{eqnarray*}
{\bf c}-\frac{1}{4}{\bf b}^{T}{\bf A}^{-1}{\bf b} &=& \text{Tr}\Bigg\{\bigg(2{\bm \Sigma}_{1}-{\bm \Sigma}_{2}\bigg)^{-1}\bigg({\bm \mu}_{1}{\bm \mu}_{1}^{T}+{\bm \mu}_{2}{\bm \mu}_{2}^{T}-{\bm \mu}_{1}{\bm \mu}_{2}^{T}-{\bm \mu}_{2}{\bm \mu}_{1}^{T}\bigg)\Bigg\} \\
&=& \bigg({\bm \mu}_{1}-{\bm \mu}_{2}\bigg)^{T}\bigg(2{\bm \Sigma}_{1}-{\bm \Sigma}_{2}\bigg)^{-1}\bigg({\bm \mu}_{1}-{\bm \mu}_{2}\bigg), 
\end{eqnarray*}
after some rearrangements. This leads to
\[
\mathbb{E}(w^{2})=\bigg(2{}^{d}|{\bm \Sigma}_{2}^{-1}-\frac{1}{2}{\bm \Sigma}_{1}^{-1}|\bigg)^{-\frac{1}{2}}|{\bm \Sigma}_{1}|^{\frac{1}{2}}|{\bm \Sigma}_{2}|^{-1}\exp\Bigg\{({\bm \mu}_{1}-{\bm \mu}_{2})^{T}(2{\bm \Sigma}_{1}-{\bm \Sigma}_{2})^{-1}({\bm \mu}_{1}-{\bm \mu}_{2})\Bigg\}.
\]
We can avoid computing the inverses of ${\bm \Sigma}_{1}$ and ${\bm \Sigma}_{2}$
by using (\ref{eq:matrixInverse}) to obtain
\[
|{\bm \Sigma}_{2}^{-1}-\frac{1}{2}{\bm \Sigma}_{1}^{-1}|{}^{-\frac{1}{2}}=\big(2^{d}|{\bm \Sigma}_{1}||2{\bm \Sigma}_{1}-{\bm \Sigma}_{2}|^{-1}|{\bm \Sigma}_{2}|\big)^{\frac{1}{2}},
\]
so finally
\[
\mathbb{E}(w^{2})=|{\bm \Sigma}_{1}||{\bm \Sigma}_{2}|^{-\frac{1}{2}}|2{\bm \Sigma}_{1}-{\bm \Sigma}_{2}|{}^{-\frac{1}{2}}\exp\Bigg\{({\bm \mu}_{1}-{\bm \mu}_{2})^{T}(2{\bm \Sigma}_{1}-{\bm \Sigma}_{2})^{-1}({\bm \mu}_{1}-{\bm \mu}_{2})\Bigg\},
\]
which exists if $2{\bm \Sigma}_{1}-{\bm \Sigma}_{2}$ is positive definite.

\section{Mixture of warped Gaussians example details\label{sec:BanDeriv}}

The gradient and Hessian of the log-density of a general weighted mixture density
$$ 
\pi({\bf x}) = \sum_{i=1}^{r}w_{i}p_{i}({\bf x}),
$$
are
\[
\nabla\log \pi({\bf x})=\sum_{i=1}^{r}\frac{w_{i}p_{i}({\bf x})}{p({\bf x})}\nabla\log p_{i}({\bf x}),
\]
\[
\nabla^{2}\log p({\bf x})=\sum_{i=1}^{r}\frac{w_{i}p_{i}({\bf x})}{p({\bf x})}\Bigg\{\nabla^{2}\log p_{i}({\bf x})+\nabla\log p_{i}({\bf x})\nabla\log p_{i}({\bf x})^{T}\Bigg\}-\nabla\log p({\bf x})\nabla\log p({\bf x})^{T}.
\]

The mixture of Section \ref{sec:banExample} is composed of $d$-dimensional warped Gaussian densities, with
parameters $a$, $b$, $s_1$ and $s_2$.  Let $\bf z$ be a $d$-dimensional vector such that $z_{1}=x_{1}-s_{1}$, $z_{2}=x_{2}-s_{2}$ and $z_{i}=x_{i}$
for $i=3,\dots,d$. Then the entries of $\nabla\log p({\bf x})$ and $\nabla^{2}\log p({\bf x})$ can be obtained by noticing that the Jacobian of this transformation is the identity matrix and using
\[
\frac{\partial\log p({\bf z})}{\partial z_{1}}=-\frac{z_{1}}{a^{2}}-2bz_{1}\{z_{2}+b(z_{1}^{2}-a^{2})\},
\]
\[
\frac{\partial\log p({\bf z})}{\partial z_{2}}=-z_{2}-b(z_{1}^{2}-a^{2}),\;\;\;\;\frac{\partial\log p({\bf z})}{\partial z_{i}}=-z_{i},\;\;\;\text{for}\;\;\;i=3,\dots,d.
\]
and
\[
\frac{\partial^{2}\log p({\bf z})}{\partial z_{1}^{2}}=-a^{-2}-2b\{z_{2}+b(z_{1}^{2}-\sigma^{2})\}-4b^{2}z_{1}^{2},
\]
\[
\frac{\partial^{2}\log p({\bf z})}{\partial z_{1}\partial z_{2}}=\frac{\partial^{2}\log p({\bf z})}{\partial z_{2}\partial z_{1}}=-2bz_{1},\;\;\;\frac{\partial^{2}\log p({\bf z})}{\partial z_{i}\partial z_{i}}=-1,\;\;\;\text{for}\;\;\;i=2,\dots,d,
\]
with all the remaining entries of the Hessian being equal to zero. In Section \ref{sec:banExample} we used six densities with the following parameters
$$
{\bf a}=\{1,6,4,4,1,1\}, \;\;\;\; {\bf b}=\{0.2,-0.03,0.1,0.1,0.1,0.1\},
$$
$$
{\bf s}_{1}=\{0,0,7,-7,7,-7\}, \;\;\;\; {\bf s}_{2}=\{0,-5,7,7,7.5,7.5\},
$$
where, for instance, the $i$-th element of $\bf a$ is the value of $a$ use to define the $i$-th warped Gaussian. The weights of the target and of the importance mixture components are
$$
{\bf w}_\text{T} \propto \{1, 4, 2.5, 2.5, 0.5, 0.5\}, \;\;\;\;
{\bf w}_{\text{IS}} \propto \{1, 4, 2.5, 2.5\}.
$$

\section{Bayesian logistic regression details \label{app:logRegDetails}}

The gradient the log-posterior is
$$
\nabla\log\pi(\bm \theta)={\bf X}^{T}{\bf y}-\sum_{i=1}^{n}\frac{{\bf X}_{i:}^T}{1+e^{-{\bf X}_{i:}^T\bm \theta}} -  \bm \alpha \odot \bm \theta,
$$
where $\alpha_1 = 0$, $\alpha_j = \lambda$ for $j= 1, \dots, d$ and $\odot$ is the Hadamard product. The Hessian is 
$$
\nabla^2\log\pi(\bm \theta)=-\sum_{i=1}^{n}\frac{e^{{\bf X}_{i:}^T\bm \theta}}{(1+e^{{\bf X}_{i:}^T\bm \theta})^{2}}{\bf X}_{i:}{\bf X}_{i:}^{T}- (\bm \alpha \bm \alpha^T)^{\frac{1}{2}},
$$
where ${\bf X}_{i:}$ is a column vector including the elements of $i$-th row of $\bf X$.

\end{appendices}

\end{document}